# The Hundredth Anniversary of Einstein's Annus Mirabilis

**Roman Ya. Kezerashvili**

*New York City College of Technology, The City University of New York*
*300 Jay Street, Brooklyn, NY 11201*

> *"I want to know how God created this world.*
> *I am not interested in this or that phenomenon,*
> *in the spectrum of this or that element.*
> *I want to know His thoughts; the rest are details."*
>
> *Albert Einstein*

## Introduction

In Germany, the year 2000 was declared the Year of Physics to celebrate the 100$^{th}$ year anniversary of the German physicist Max Planck's announcement that by making the strange modification in the classical calculation he could explain the most puzzling phenomena studied nearly at the end of the nineteenth century: the spectral distribution of blackbody radiation. Planck first found an empirical function that fit the data by making the unusual assumption that the energy emitted and absorbed by the blackbody was not continuous but was instead emitted and absorbed in discrete packets or *quanta*. Plank found that the size of an energy quantum is proportional to the frequency of the radiation:

$$E = hf\,, \qquad (1)$$

where *h* is the proportionality constant now known as Planck's constant and *f* is the frequency of radiation. The value of *h* was determined by Planck by fitting his function to the experimentally obtained data. He was unable to fit the constant into framework of classical physics.

The Year of Physics inspired the European Physical Society to suggest that 2005 be declared the World Year of Physics. This has been taken up by the International Union of Pure and Applied Physics, by UNESCO, by the Congress of the United States, and by the General Assembly of the United Nations, which declared 2005 the World Year of Physics.

## Why 2005?

The year 2005 marks the hundredth anniversary of the pioneering contributions of Albert Einstein to modern physics in 1905 and commemorates his great ideas, and his influence on life in the 20th century. 1905 is the "Annus Mirabilis" – Einstein's special or "Miracle Year".

In 1905, Einstein submitted for publication five groundbreaking fundamental papers, all in a few months, which changed our understanding of the world. The first paper claimed that light must sometimes behave like a stream of particles with discrete energies, "quanta." The other two papers offered an experimental test for the theory of heat and proof of the existence of atoms – the reality of the discontinuous, atomic nature of matter. The fourth paper addressed the central puzzle for physicists of the day – the connection between electromagnetic theory and ordinary motion – and solved it using the "principle of relativity." The fifth showed that mass and energy are two parts of the same thing, mass-energy. These papers revolutionized physics, changed history and the face of our civilization.

The resolution of the House of Representatives states that Congress " encourages the American people to observe the World Year of Physics as a special occasion for giving impetus to education and research in physics as well as to the public understanding of physics", and it acknowledges the contribution of physics to "knowledge, civilization, and culture".

**Quantum Idea**

The fundamental importance of Planck's assumption of energy quantization, implied by equation (1), was considered as a mathematical trick and was not generally appreciated until Einstein applied similar idea to explain the photoelectric effect and suggested that quantization is a fundamental property of electromagnetic radiation. On March 17, 1905, *Annalen der Physik*, the leading German physics journal, received Einstein's paper [1] *"On a heuristic viewpoint concerning the generation and transformation of light",* where he used Planck's quantum hypothesis of energy quantization to explain the photoelectric effect and gave a quantitative theory of this phenomenon.

In the early 1700s Isaac Newton, who first became famous for his experiments with light, proposed that light was made up of tiny particles – corpuscles. Then a hundred years later the wave nature of light was demonstrated by Thomas Young in his famous double-slit experiment. This wave view was reinforced in 1862 by James Maxwell's finding that light is carried in the oscillating electric and magnetic fields of electromagnetic waves. Numerous experiments on the interference, diffraction, and scattering of light had confirmed the wave nature of light. Heinrich Hertz confirmed the wave view of light experimentally in 1887, eight years after Maxwell's death. We can well appreciate the shock and disbelief when Einstein in 1905 challenged the wave theory of light. Einstein stated that, in its interaction with matter, light is confined not in continuous waves but in tiny particles of energy called photons. So physics of light had come full circle in its view of its nature: particle to wave and back to particle.

Let's look at Einstein's particle model of light, which explained the photoelectric effect – a phenomenon that puzzled scientists for many years since Heinrich Hertz had discovered it in 1887. When Heinrich Hertz performed his experiments to prove the existence of electromagnetic waves, he accidentally found that when light falls fell on a certain metallic surface, electrons are ejected from the surface. This is the photoelectric effect, used in electric eyes, light meters, and motion-picture sound tracks. Careful investigations toward the end of the nineteenth century proved that the photoelectric effect occurs with different materials but only if the wavelength of light is short enough. Ultraviolet and violet light strikes the surface with sufficient energy to knock electrons from the surface of a metals but lower-frequency light does not – even if the lower-frequency light is very bright. Thus, the photoelectric effect has some threshold wavelength, which is specific to the material. Also, ejection of electrons depends only on the frequency of light, and the higher the frequency of light used, the greater the kinetic energy of the ejected electrons. The fact that light of large wavelengths has no effect at all even if it is extremely intensive appeared especially mysterious for the scientists.

Albert Einstein's explanation was that electrons in the metals are bombarded by the "particles of light" – photons and the energy of each photon is proportional to the frequency of the light. There is a certain minimum amount of energy (dependent on the material), which is necessary to remove an electron from the surface of a metal or another solid body. If the energy of a photon is bigger than this value, the electron can be emitted. Thus, Einstein assumed that a photon with energy given by equation (1) struck an electron on the metallic surface and the electron then absorbed its entire quantum of energy. A portion of this energy is used by the electron to break away from the metallic surface, and the rest shows up as the kinetic energy of the electron. From this explanation the following equation results:

$$KE_{max} = hf - W . \qquad (2)$$

Equation (2) is known as Einstein's photoelectric equation. In equation (2) the quantity $W$ is called the work function and it is equal to minimum energy of the photons when the photoelectric effect occurs, $KE_{max}$ is maximum kinetic energy of the emitted electron, $h$ is the Planck's constant and $f$ is the frequency of radiation.

Einstein's photoelectric equation was a bold prediction, for at the time it was made there was no evidence that Planck's constant had any applicability outside of blackbody radiation. Experimental verification of Einstein's theory was quite difficult. Careful experiments by Robert Millikan, reported first in 1914 and then in more detail in 1916, showed that Einstein's equation was correct and the measurements of Planck's constant $h$ agreed with the value founded by Planck. In 1923, Robert Millikan was awarded a Nobel Prize for his work on the elementary charge of electricity and on the photoelectric effect.

In essence, Planck had discovered the quantum structure of electromagnetic radiation. But Planck himself did not see it that way; he saw the new assumption merely as a mathematical trick to obtain the right answer. Its significance remained for him a mystery. Planck's formula (1) had survived for half a decade. However, it lacked a physical basis, which could be taken seriously by anyone. Einstein went far beyond this, to the next step, to use Planck's quantum hypotheses, but from an entirely innovative approach. Light behaves not as continuous waves but as discontinuous, individual particles – photons and the energy of such a photon is proportional to the frequency of the light.

**Existence of Atoms**

In May 11 of 1905**,** *Annalen der Physik*, the world's leading physics journal, received Einstein's paper [2] **"***On the motion of particles suspended in a stationary fluid, as demanded by the molecular kinetic theory of heat"*. This was Einstein's first paper on the Brownian motion, named after the Scottish botanist Robert Brown, who is credited with its discovery in 1827. While he observed tiny pollen grains suspended in water under his microscope, Brown noticed that tiny grains moved about a tortuous path, even though the water appeared to be perfect still. The atomic theory easily explains Brownian motion if the further reasonable assumption is made that atoms of any substance are continually in motion. Then Brown's tiny pollen grains are jostled about by the vigorous barrage of rapidly moving molecules of water. Albert Einstein used a statistical approach and examined Brownian motion from a theoretical point of view, and derived predictions for the displacement of microscopic particles suspended in a fluid. If tiny but visible particles were suspended in a liquid, he said, the irregular bombardment by the liquid's invisible atoms should cause the suspended particles to carry out a random jittering dance. In two other papers [3,4] (the article [3] "*A new determination of molecular dimensions"*, also was Einstein's doctoral thesis submitted to the University of Zurich) published in 1906, Einstein extends his analysis of the Brownian motion to the rotational motion of suspended particles and calculated from experimental data the approximate size and mass of molecules. His calculations showed that the diameter of a typical atom is about $10^{-10}$m. For many years this was his most cited study. Thus, among the things Einstein did in 1905 were the proof that molecules (and, therefore, by extension, the atoms of which they are composed) actually exist, and the prediction of their size. A French scientist, Jean Baptiste Perrin, conducted a series of experiments that confirmed Einstein's predictions for Brownian motion and therefore, the reality of the discontinuous nature of matter. His work earned him his own Nobel Prize in Physics in 1926.

**Theory of Special Relativity**

The most revolutionary conceptual change arrived in 1905, 30 June, when Einstein submitted his first paper [5] on Special Relativity "*On the electrodynamics of moving bodies"*. First of all just what was meant by the word *relativity*? The term refers to the relative motion of two reference frames. The special theory of relativity concentrates on the relationship between events and physical quantities specified in different inertial reference frames.

The theory of special relativity can be derived from two postulates proposed by Einstein, one rooted in experiment, the other stemming from aesthetic, even natural philosophy as well as experiment, and argument about the apparent experimental equivalence of all inertial reference frames. The first postulate states

- All fundamental laws of physics must be the same in all inertial reference frames.

The second postulate encompasses all measurements, early and modern, of the speed of light and the prediction of the speed of light by Maxwell's electromagnetic theory and states

- The speed of light in a vacuum has the same numerical value *c* when measured in any inertial reference frame, independent of the motion of the source of light and/or observer.

The second postulate contradicts our intuitive ideas about relative velocities. It tells us that the speed of light in a vacuum is always the same, no matter what the speed of the observer or the source. Thus, a person traveling toward or away from a source of light will measure the same speed for that light as someone at rest with respect to the source. This conflicts with our every day experience: we would expect to have to add in the velocity of the observer. Regardless of the confounding nature of the second postulate, it reflects the way nature is, and physics must take nature on its own terms.

Relativity in Newtonian physics involves certain unprovable assumptions that make sense from everyday experience. It is assumed that the lengths of the object are the same in one frame as in another, and time passes at the same rate in different reference frames. In Newtonian physics, subsequently space and time intervals are considered to be absolute, and their measurement does not change from one reference frame to another. Einstein's relativity changes our understanding of space and time. They are not absolute. We cannot speak meaningfully about space without implying time. Things exist in space-time. Einstein's postulates change concepts about space and time and are part of a large picture, a revolutionary one that predicts that motion through space causes time to slow down, that objects in motion are shorter and mass is actually congealed energy.

Based on these postulates Einstein derived transformations that make Maxwell's electromagnetic equations invariant in all inertial reference frames. These transformations known today as Lorentz transformation equations were first proposed in a slightly different form, by Lorentz in 1904, to explain "the null result" of the Michelson-Morley experiment. Interestingly, he justified the transformation on what was eventually discovered to be a fallacious hypothesis. Contradicting Lorentz, Einstein introduced the Lorentz transformation equations to account for the peculiar constancy of the speed of light, an invariance that violates the Galilean transformation equations. According to Einstein the laws of physics are invariant to the Lorentz transformationequations between inertial frames.

One of the important consequences of the theory of relativity is that we can no longer regard time as an absolute quantity. No one doubts that time flows onward and never turns back. But, the time interval between two events, and even whether two events are simultaneous, depends on the reference frame. The time interval between two events that occur at the same reference frame is always less than the time interval between the same events that were measured in another reference frame in which the event occurs at a different place. This is a general result of the special theory of relativity, and is known as time dilation. This relationship between the time intervals is given as

$$\tau = \frac{\tau_0}{\sqrt{1 - v^2/c^2}}.  \qquad (3)$$

In equation (3), $\tau_0$ is the rest time interval because it is measured in the frame in which the clock is at rest, $\tau$ is the time interval measured by the clock in the frame which is moving with the speed *v*. Thus, from equation (3) it follows that the moving clock runs slower than the identical clock at rest. This is called a time dilation effect. For example, today common electronic devices on the Global Positioning System satellites have to take dilation effect into account in order to function properly.

Not only are time intervals different in different reference frames. Space intervals – lengths and distances – are different as well, according to the special theory of relativity. That is

$$L = L_0 \sqrt{1 - v^2/c^2} ,  \qquad (4)$$

where $v$ is the relative velocity between observed and observer, $c$ is the speed of light, $L$ is the measured length of the moving object, and $L_0$ is the measured length of the object at rest. This length contraction was first proposed by George FitzGerald and mathematically expressed by Hendrik Lorentz before Einstein's paper was published. Whereas these physicists hypothesized that matter contracts in order to explain "the null result" of the Michelson-Morley experiment, Einstein saw that what contract is space itself. Nevertheless, because Einstein's formula is the same as Lorentz's, we call the effect the Lorentz – FitzGerald contraction. This is a general result of the special theory of relativity and applies to length of object as well as to distance. The result can be summarized by saying that the length of an object in motion with respect to an observer is less than its length when measured by an observer who is at rest with respect to the object. This contraction occurs only in the direction of the relative motion.

Einstein linked not only space and time but also mass and energy. A piece of matter, even at rest and not interacting with anything else, has an energy being. This is called its rest energy. Here we have Einstein's famous formula, which shows how the amount of energy $E$ is related to the amount of mass $m$

$$E = mc^2 . \qquad (5)$$

This is the most celebrated equation of the 20th century, which brought with it the dawning of the nuclear age. This formula mathematically relates the concepts of energy and mass through square of the speed of light, $c^2$ – the conversion factor between energy units and mass units. This idea was published in 1905 in a paper[6] entitled *"Does the inertia of a body depend on its energy content?"* But if this idea is to have any meaning from a practical point of view, then mass ought to be convertible to energy and vice versa. That is, if mass is just one form of energy, then it should be convertible to other form of energy as other types of energy are interconvertible. Einstein suggested that this might be possible, and indeed changes of mass to the other forms of energy, and vice versa, have been experimentally confirmed countless times. The interconversion of mass and energy is most easily detected in nuclear and elementary particle physics. On a large scale, the radiant energy we receive from the Sun is an example of equation (5). When gravitation crunches a mass of the Sun and ignites a thermonuclear fusion, the energy that emerges is accompanied by the corresponding lowering of mass – but only a tiny bit. The helium nucleus produced by the fusion of a pair of deuterium nuclei is about one-thousandth less massive than two deuterium nuclei. Sunlight, then, is this small amount of mass transformed by thermonuclear fusion into radiant energy. Sun's mass is constantly decreasing as it radiates electromagnetic energy. The energy produced in nuclear power plants is a result of the loss in the rest mass of the uranium fuel as it undergoes the process called fission. A tremendous amount of energy is released in the fission reaction because the mass of uranium is considerably greater than the total mass of the fission fragments plus emitted neutrons. In chemical reactions where heat is gained or lost, the masses of the reactants and the products will be different. Even when water is heated on the stove, the mass of the water increases very slightly.

All the above demonstrated the triumph of Einstein's "Annus Mirabilis". He showed that atoms are real, an idea that was still controversial at the time, presented his special theory of relativity, and put quantum theory on its feet. In 1921 Albert Einstein was awarded a Nobel Prize "for his services to Theoretical Physics, and especially for his discovery of the law of the photoelectric effect". Albert Einstein served Theoretical Physics until the end of his life in 1955 (making 2005 also the 50th anniversary of his death). His remarkable results in the year 1905 so dramatically transformed our understanding and ideas about the microworld and cosmos that after one hundred years it can fairly be said that because of his remarkable achievements humanity occupies a completely different universe from the one that was imagined only three generations earlier.